\documentclass{aa}  

\usepackage{xcolor}
\usepackage{natbib}
\usepackage{graphicx}
\usepackage{txfonts}
\usepackage[
  breaklinks = true,
   colorlinks = true,
  urlcolor   = blue,
  citecolor  = blue,
  linkcolor  = blue,
]{hyperref}

\def\kms{\hbox{km$\;$s$^{-1}$}}
\def\Halpha{\mbox{H\hspace{0.1ex}$\alpha$}}
\def\Hbeta{\mbox{H\hspace{0.1ex}$\beta$ }}
\newcommand{\kmeans}{\textit{k}-means}
\defcitealias{2025A&A...693A...8T}{Paper~I}

\begin{document} 

   \title{Fine details in solar flare ribbons}
   \subtitle{Statistical insights from observations with the Swedish 1-m Solar Telescope}

    \author{
    Jonas Thoen Faber \inst{1,2}
    \and
    Reetika Joshi \inst{3,4,1,2}
    \and
    Luc Rouppe van der Voort \inst{1,2}
    \and
    Sven Wedemeyer \inst{1,2}
    \and
    Eilif Sommer Øyre \inst{1,2}
    \and
    Ignasi J. Soler Poquet \inst{1,2}
    \and
    Aline Rangøy Brunvoll \inst{1,2}
    }

    \institute{Rosseland Centre for Solar Physics, University of Oslo, PO Box 1029, Blindern 0315 Oslo, Norway;\\
    \email{j.t.faber@astro.uio.no}
    \and
    Institute of Theoretical Astrophysics, University of Oslo, PO Box 1029, Blindern 0315 Oslo, Norway
    \and
    NASA Goddard Space Flight Center, Heliophysics Science Division, Code 671, Greenbelt, MD 20771, USA
    \and
    Department of Physics and Astronomy, George Mason University, Fairfax, VA 22030, USA
    }
    \date{Received  / Accepted  }

 
  \abstract
   {
   Flare ribbons serve as chromospheric footprints of energy deposition resulting from particle acceleration during magnetic reconnection. Their fine-scale structure provides a valuable tool for probing the dynamics of the flare reconnection process.
   }
   {
   Our goal is to investigate the fine-scale structure of flare ribbons through multiple observations of flares, utilising data obtained from the Atmospheric Imaging Assembly (AIA) and the Swedish 1-m Solar Telescope (SST).
   }
   {
   The aligned AIA and SST datasets for the three solar flares were used to examine their overall morphology. The SST datasets were specifically used to identify fine-scale structures within the flare ribbons. For spectroscopic analysis of these fine structures, we applied machine-learning methods (\kmeans\ clustering) and Gaussian fitting. 
   }
   {
   Using \kmeans, we identified elongated features in the flare ribbons, termed as ``riblets'', which are short-lived and jet-like small-scale structures that extend as plasma columns from the flare ribbons. Riblets are more prominent near the solar limb and represent the ribbon front.
   Riblet widths are consistent across observations, ranging from 110--310~km (0\farcs 15--0\farcs 41), while vertical lengths span 620--1220~km (0\farcs 83--1\farcs 66), with a potential maximum of 2000~km (2\farcs 67), after accounting for projection effects. Detailed \Hbeta spectral analysis reveals that riblets exhibit a single, redshifted emission component, with velocities of 16--21~\kms, independent of viewing angle.
   }
   {
   Our high-resolution observations of the three flare ribbons show that they are not continuous structures, but are composed of vertically extended, fine-scale substructures. These irregular features indicate that the reconnection region is not a smooth, laminar current sheet, but rather a fragmented zone filled with magnetic islands (blobs or riblets), consistent with the theory of patchy reconnection within the coronal current sheet.
   }


   \maketitle


\section{Introduction}
\label{sec:introduction}

Solar flares rank among the most energetic phenomena originating from the Sun. These are observable across the entire electromagnetic spectrum, from $\gamma$- and X-rays to radio wavelengths, often exhibiting ribbon-like structures known as ``flare ribbons'' \citep{Acton1982, Masuda1994, 2008LRSP....5....1B, Shibata2011, Yang2015, Devi2020, Joshi2025}. The formation of flares and these ribbons has been extensively explained by theoretical models, starting with the 2D standard model \citep[CSHKP,][]{1964NASSP..50..451C, 1966Natur.211..695S, 1974SoPh...34..323H, 1976SoPh...50...85K}, which was later extended to three dimensions \citep{2012A&A...543A.110A}. In the CSHKP model, magnetic reconnection occurs within a vertical current sheet formed between oppositely directed magnetic field lines. Energy released at the coronal reconnection site is transported downward, heating the flare ribbon structures in the lower solar atmosphere. Hence, flare ribbons are believed to represent the chromospheric footpoints of magnetic reconnection in the solar corona \citep{Schmieder:1987ApJ...317..956S,  2008LRSP....5....1B, Chandra2009, 2011SSRv..159...19F, 2017A&A...601A..26Z, 2024A&A...687A.172J}.
Due to the difficulty of observing the reconnection process, even with current high-resolution observations, ribbons are interpreted as a consequence of the magnetic field reconfiguration.

There have been observational efforts to probe the fine structures in these flare ribbons in detail \citep{Warren2000, 2014ApJ...788L..18S, 2016NatSR...624319J, 2021ApJ...922..117F, 2025A&A...693A...8T, 2025arXiv250701169S, 2025ApJ...989..183Y}. In addition to these observational studies, \citet{Pietrow2024} conducted an extensive analysis of fine-scale structures using spectropolarimetric observations of an X-class flare.
Numerical models have been developed to detail the dynamics of the plasma in the solar atmosphere, aiming to explain the occurrence of magnetic reconnection in the corona and probe the structural signatures in the layers below into which a substantial portion of the released energy is transported. For instance, \citet{2021ApJ...920..102W} analysed a static 3D magnetic field configuration of a flux rope and suggested that fine-scale structures formed in flare ribbons are due to oblique tearing. Very recently, \citet{2025arXiv250400913D} applied the concept of ``field-line length map'' on a 3D simulation to analyse the length of magnetic field lines mapped to the surface. They conclude that downward propagating plasmoids formed in the current sheet are the source of small-scale features along the ribbon fronts. While these studies analysed the configuration of the solar atmosphere of a flare, explaining the energy transport from the reconnection site down to the footpoints is not included \citep[see, e.g.][]{Kerr2021, 2013A&A...558A..76R}.

In our previous study \citep[][hereafter \citetalias{2025A&A...693A...8T}]{2025A&A...693A...8T}, we analysed the ribbon fine structures of a C2.4 class flare (SOL2022-06-26T08:12) located near disk center (DC). The analysis included co-observations from the Atmospheric Imaging Assembly \citep[AIA,][]{2012SoPh..275...17L} on board the Solar Dynamics Observatory \citep[SDO,][]{2012SoPh..275....3P}, the Swedish 1-m Solar Telescope \citep[SST,][]{2003SPIE.4853..341S}, and the Interface Region Imaging Spectrograph \citep[IRIS,][]{2014SoPh..289.2733D}. These events, which could be recognised as flare kernels in \ion{Si}{IV}~1400~\AA\ \citep[see, e.g.][]{2022ApJ...934...80L}, were referred to as ``blobs'' due to their spatial periodicity and their circular structures in \ion{Ca}{II}~8542 and \Hbeta wing images. To better understand the processes related to the formation of small-scale structures in flare ribbons, it is crucial to investigate these structures from multiple viewing angles, which provide different perspectives of the events. In the study presented here, we aim to identify such structures in flare ribbons with varying strengths and viewing angles, and to conduct a statistical analysis of these fine-scale structures. In this analysis, we refer to these bright and extended fine-structure in ribbons instead as ``riblets'', following \citet{2025arXiv250701169S} who analysed SST \Halpha\ observations of an X1.5 flare occurring on 10 June 2014 near the limb \citep[also see Fig.~7 in ][]{Pietrow2024}. Riblets are sudden and bright small-scale features located along the length of the ribbons, mainly seen in the red wings of chromospheric lines. Their lifetimes are typically on the order of several seconds before merging back into the ribbon structure. Different flares generated from seemingly similar morphologies add comparable diagnostics to the analysis. The findings in \citetalias{2025A&A...693A...8T} showed that H$\beta$ observations were optimal for studying the blobs, which motivates the use of the same diagnostic channel.

The paper is structured as follows. We present the three flare observations in Sect.~\ref{sec:observations} and provide context in terms of flare morphology. Identification of the fine structures using machine learning is explained in Sect.~\ref{sec:methods}. Additionally, we explain the parameters extracted when performing a line fit to the relevant spectral lines. The results from an imaging and spectral view are presented in Sect.~\ref{sec:results}. We discuss and conclude our findings in Sect.~\ref{sec:discussion} and Sect.~\ref{sec:conclusions}, respectively.

\begin{figure*}[ht!]
    \sidecaption
    \includegraphics[width=\textwidth]{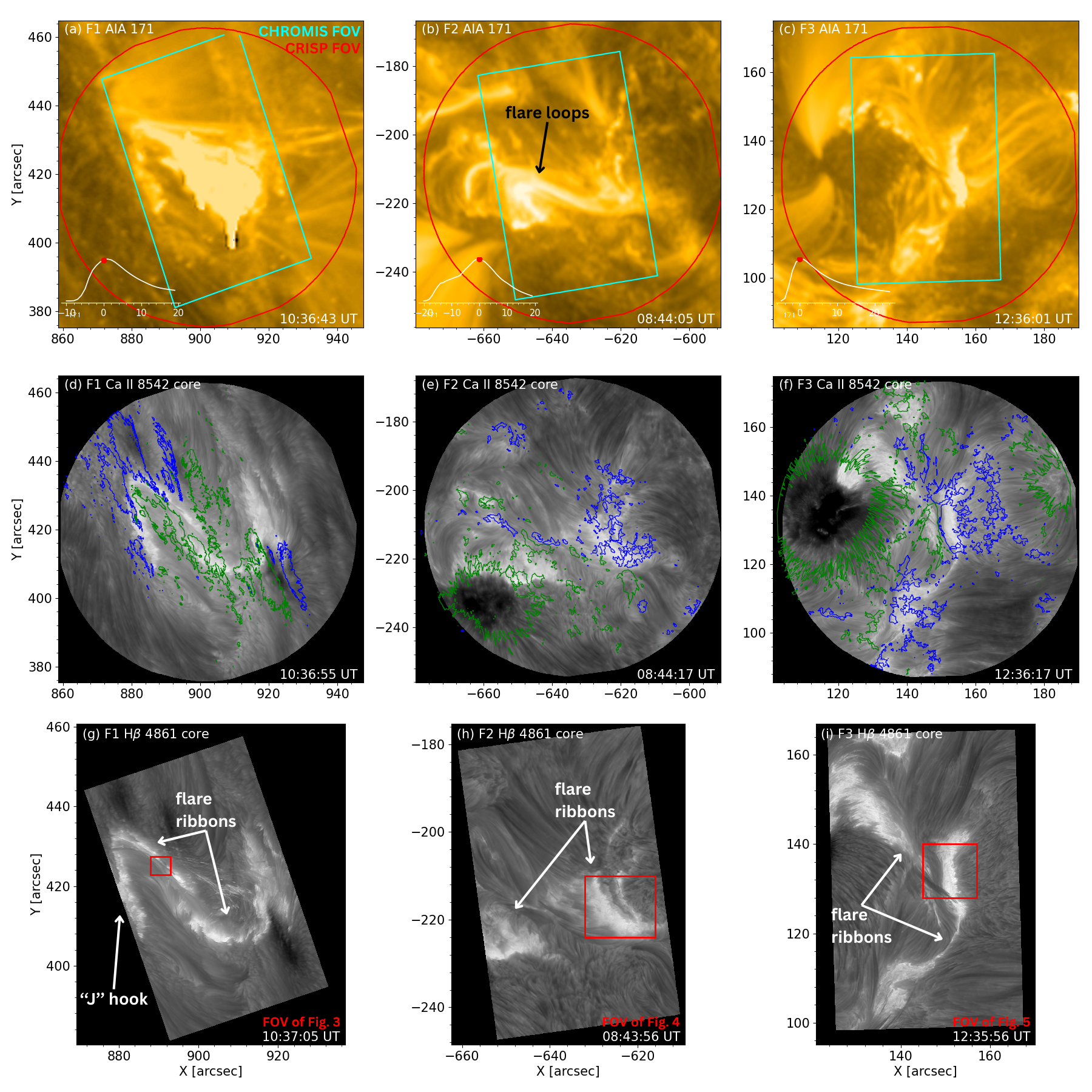}
    \centering
    \caption{Context of the ARs that generated the flares. The columns from left to right show the M4.6, the C8.3, and the M1.8 flares, respectively. The rows from top to bottom show the AIA 171~\AA, \ion{Ca}{II}~8542~\AA\ core and H$\beta$ core channels, respectively. Each panel show the flare near the GOES peak time and the timestamps are shown in the lower-right corners. The GOES X-ray plot is added in the lower left corner in the upper row where the x-axis is in minutes relative to the image. The red and cyan contour highlights the FOV of CRISP and CHROMIS, respectively. The green and blue contours in the middle row show the CRISP magnetogram at $\pm$ 500 G. The red boxes in the lower row correspond to the FOVs in Figs.~\ref{fig:F1_kmeans_and_Gauss_maps}--\ref{fig:F3_kmeans_and_Gauss_maps} respectively. An animation of this figure is available online.
    }
    \label{fig:ALL_aia171_context}
\end{figure*}

\begin{figure*}[ht!]
    \sidecaption
    \includegraphics[width=\textwidth]{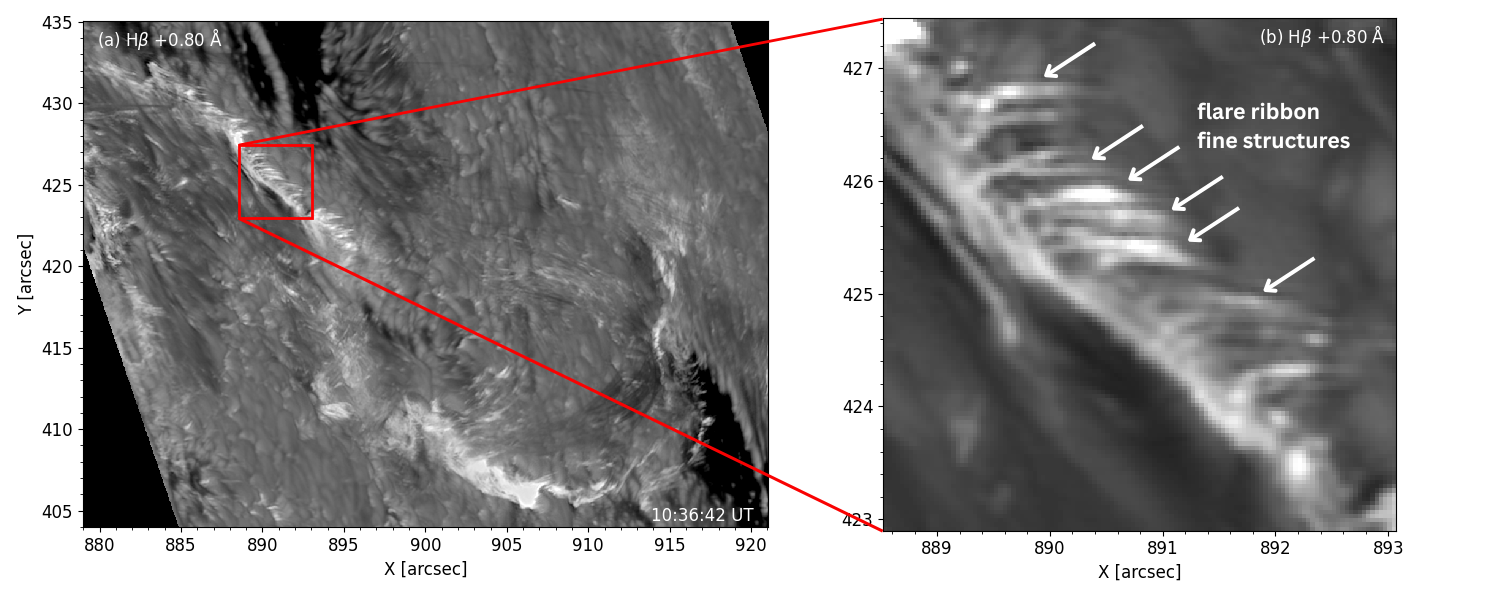}
    \centering
    \caption{
    Fine-scale structures or riblets in a flare ribbon located near the western limb in a complex photospheric magnetic field configuration (flare F1). The right panel is a smaller FOV, as highlighted by the red rectangle in the left panel, revealing the fine-scale structures in a ribbon from a side-view. The colourmaps in the left and right panels are in logarithmic and linear scales, respectively.
    }
    \label{fig:F1_riblet_context}
\end{figure*}


\section{Observations}
\label{sec:observations}

\subsection{Flare selection}

To complement the analysis in \citetalias{2025A&A...693A...8T}, we selected three different data sets from our flare observations from the 2023 and 2024 SST observing seasons. These data sets were selected for their consistent high seeing quality and coverage of the Geostationary Operational Environmental Satellite (GOES) X-ray flare peak. The three flares were of medium strength, having GOES class C8.3, M1.8, and M4.6. The three flares had varying positions on the disk: close to the limb, intermediate, and close to the disk centre. These flares are labelled as F1, F2 and F3, respectively. The different viewing angles provide varying perspectives on the fine structures in a ribbon. Table~\ref{tab:context} provides details of the different data sets, and Fig.~\ref{fig:ALL_aia171_context} presents context. These high-resolution observations provide additional diagnostics to the flare ribbon blobs that we analysed in \citetalias{2025A&A...693A...8T}. A context image of such fine-scale structures captured from a side view is shown in Fig.~\ref{fig:F1_riblet_context}.

\subsection{Flare morphology}

The three flares occurred in different active regions. 
Only AR 13376 (F1) had two sunspots with opposite polarities and developed the strongest flare in the list. While a single sunspot was evident in the two other ARs, strong magnetic fields were evident in the form of prominent flux concentrations in SST \ion{Fe}{I}~6173~\AA\ magnetograms. Each flare was associated with an erupting filament seen in AIA channels with one end seemingly connected to a sunspot penumbra. Two ribbons connected above strong patches of opposite-directed polarity regions in the photosphere are evident for all three flares. For the F1 and F3 flares, the ribbons of opposite polarities were seemingly anti-parallel, resembling the characteristic ``J-shape`` ribbon pair, indicated by the white arrows in Fig.~\ref{fig:ALL_aia171_context}g and Fig.~\ref{fig:ALL_aia171_context}i, respectively. The underlying opposite-directed magnetic field is visible in Fig.~\ref{fig:ALL_aia171_context}d and Fig.~\ref{fig:ALL_aia171_context}f by the green and blue contours. The F2 flare exhibits a pair of ribbons, but an anti-parallel configuration is not evident. The straight parts of the ribbons are nearly perpendicular relative to one another. Additionally, the ribbon closer to the sunspot lacks a well-defined hook. The evolution of each flare is presented in Fig.~\ref{fig:ALL_aia171_context} (see also the associated animation in the online material). In the AIA~171~\AA\ coronal channel, flare loops connecting the opposite polarity ribbons are visible (indicated by the black arrow in Fig.~\ref{fig:ALL_aia171_context}b). The flare loops from the F1 flare in Fig.~\ref{fig:ALL_aia171_context}a saturate the image.

\subsection{Observing programs and data reduction}

The three flare events were observed with the CRisp Imaging SpectroPolarimeter \citep[CRISP,][]{2008ApJ...689L..69S} and the CHROMospheric Imaging Spectrometer 
\citep[CHROMIS,][]{2017psio.confE..85S, 2025arXiv250605143S} 
installed at the Swedish 1-m Solar Telescope \citep[SST,][]{2003SPIE.4853..341S}.
CRISP was running a program similar to what is described in \citetalias{2025A&A...693A...8T}, sampling the \ion{Fe}{i}~6173~\AA\ and \ion{Ca}{ii}~8542~\AA\ lines in spectropolarimetric mode.
The F1 and F2 flares include the \Halpha\ line in spectral imaging mode. 
The field of view (FOV) of CRISP has a diameter 87\arcsec, which is larger than for the observations in \citetalias{2025A&A...693A...8T}. 
Here, we concentrate on the CHROMIS observations which sampled the \Hbeta\ line in 29~line positions between $\pm2.7$~\AA. The sampling steps were 0.1~\AA\ in the core region between $\pm$1~\AA, and coarser (varying between 0.15--0.8~\AA) in the wings to avoid spectral line blends. 
The cadence was about 11~s. 
CHROMIS had a pixel scale of about 0\farcs038 and a FOV of about 72\arcsec $\times$ 48\arcsec. 
The diffraction limit $\lambda/D$ of the SST with its $D=0.97$~m aperture is about 0\farcs1 at $\lambda=4861$~\AA. 
The data was processed following the standard SSTRED data reduction pipeline
\citep{2015A&A...573A..40D, 
2021A&A...653A..68L} 
which includes Multi-Object Multi-Frame Blind Deconvolution 
\citep[MOMFBD,][]{2005SoPh..228..191V} 
image restoration. 
We performed an AIA-to-SST alignment to add context to the AR associated with the respective flares.
The AIA~171~\AA\ channel provides insights into the magnetic configuration in the corona during the evolution of the flares.

\begin{table*}
    \centering 
    \caption{Flare list. Characteristics of each flare in study.}
    \begin{tabular}{c c c c c c c c c c c} 
    
    \hline\hline 
     & NOAA & GOES &  & Start & Peak & Viewing & CHROMIS & CRISP & Flare \\
    
    Solar Target ID & AR & Class & Location & Time & Time & Angle $\mu$ & Start & Start & ID \\
    
    \hline 
    
    SOL2023-07-26T10:37 & 13376 & M4.6 & N21W70 & 10:17 & 10:37 & 0.35 & 10:36 & 10:36 & F1 \\ 

    SOL2023-07-26T08:44 & 13380 & C8.3 & S10E36 & 08:24 & 08:44 & 0.79 & 08:29 & 08:28 & F2 \\ 

    SOL2024-09-11T12:36 & 13814 & M1.8 & N15W08 & 12:27 & 12:36 & 0.98 & 12:30 & 12:30 & F3 \\ 
    
    \hline 
    \label{tab:context}
    \end{tabular}
\end{table*}

\begin{figure*}[t!]
    \sidecaption
    \includegraphics[width=\textwidth]{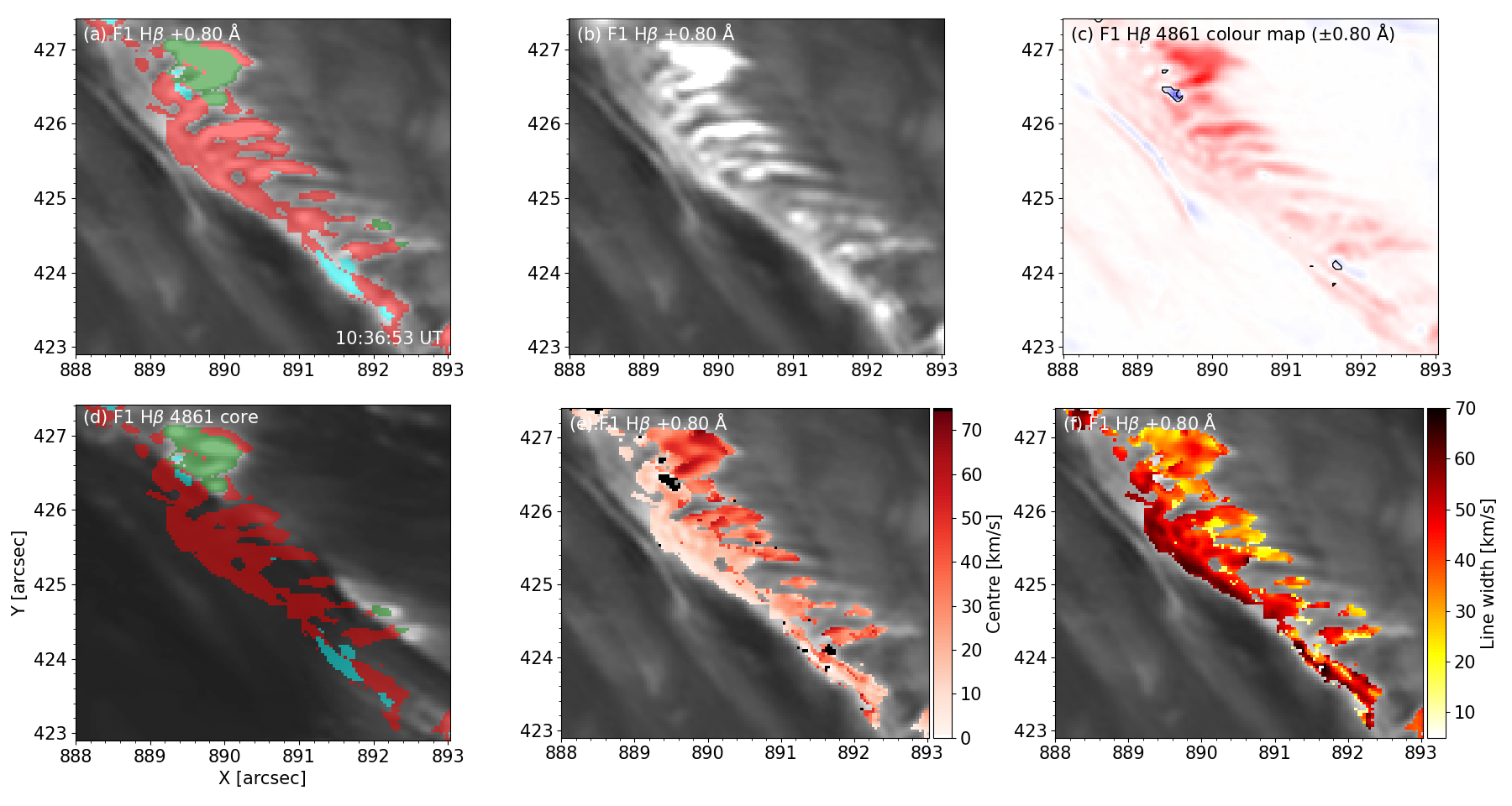}
    \centering
    \caption{
    Riblets along the F1 ribbon. All panels show the same FOV, highlighting the central parts of the eastern ribbon. 
    All panels except panel~(c)--(d) show images in the H$\beta$~+0.8~\AA\ channel. Panel~(c) shows the wing-subtracted colourmap at H$\beta \pm$~0.8~\AA. Panel~(a) shows the ribbon with identified riblets overplotted in green, red or cyan. Green pixels represent single-peaked RPs, red pixels represent double-peaked RPs with a stronger red peak, and cyan pixels represent near-symmetric double-peaked RPs. Similar is shown in panel~(d) overplotted on the H$\beta$~core image. The resulting Doppler shifts and profile widths obtained from fitting the pixels are shown in panel~(e)--(f), respectively. The black pixels in panel~(e) mask the pixels where blue shifts were estimated. The contours of these pixels are added in panel~(c). An animation related to this figure is available online.
    }
    \label{fig:F1_kmeans_and_Gauss_maps}
\end{figure*}


\section{Methods}
\label{sec:methods}

\subsection{\kmeans}

To tackle the vast amount of data points, machine learning techniques are employed. We used \kmeans\ clustering to effectively group pixels into \textit{k} clusters that represent a spectral profile with a unique shape \citep{Everitt_1972}. These groups are referred to as representative profiles (RPs). An optimised initialisation of each cluster centroid is based on the idea of \cite{10.5555/1283383.1283494}, where new centroids are drawn far away from previous centroids in terms of Euclidean distance. This method has previously been proven to be a helpful tool for grouping large numbers of data points from different solar events into representative groups \citep[see, e.g.][]{2022A&A...664A..72J, 2023ApJ...956...85T, 2024ApJS..271...24S, 2025ApJ...986..124C}. We utilised the open-source Scikit-learn package \citep{2011JMLR...12.2825P}. The learning procedure iteratively modifies the shape of the RP based on the number of pixels assigned to that particular RP cluster. We found that \textit{k}=100 for the F2 flare and \textit{k}=120 for the F1 and F3 flares were optimal for our analysis. A detailed explanation on determining \textit{k} is provided in Sect.~\ref{app:number_of_clusters}.

Clustering all pixels from a dataset is excessive since the analysis is restricted to the fine-scale features in flare ribbons. Additionally, including all pixels for clustering could potentially omit rare but important RPs. For these reasons, a three-step masking routine was developed to confine the analysis to the ribbon areas. In the first step, we applied an intensity threshold to the integration over a few selected wavelengths in the line core region. In the second step, we manually removed pixel areas that were brighter than the intensity threshold but were located clearly outside the ribbon. This step was aided by the aligned AIA images. In the third step, we applied morphological erosion and dilation operations to slightly grow the masked areas to make sure all relevant pixels were included.

\subsection{Identify fine structures in ribbons}
\label{subsec:identifying_riblets}

We aim to identify fine structures in a ribbon that appear close in time to the GOES X-ray peaks. The identification process is performed using H$\beta$~+0.8~\AA\ wing images. In \citetalias{2025A&A...693A...8T}, we referred to these features as ``blobs'' due to their spatial periodicity and circular structures. Here, we refer to these features as ``riblets''. An example of riblets is presented in Fig.~\ref{fig:F1_riblet_context}. Pixels in H$\beta$~+0.8\AA\ that are part of such structures were labelled, and the RPs associated with these pixels were tagged accordingly. We performed the identification step at a smaller FOV for multiple frames to ensure all relevant RPs were selected. If an RP was associated with an excessive portion of the ribbon that is not considered a riblet, we disregarded that particular RP from further analysis.

The RPs associated with riblets revealed that their spectral shapes are diverse. We organised the RPs in three subgroups based on their spectral properties. One subgroup is associated with RPs that consists of a single emission peak. The other two subgroups are associated with profiles with double emission peaks. They were segregated based on whether the red peak is stronger or if both peaks are equally strong. The subgroups allow us to distinguish between the different spectral characteristics along the flare ribbon.

\subsection{Emission profile fitting}

Flaring profiles are often recognised by their emission signatures that can be approximated to a Gaussian profile. All pixels associated with the blobs detected in \citetalias{2025A&A...693A...8T} show the H$\beta$ line in emission. In this work, we use the labelled pixels from the \kmeans\ clustering and apply Gaussian fitting on individual pixels associated with a riblet to extract their spectral properties. The profiles are fitted to the following function.
\begin{equation}
    I(\lambda) = A e^{-\frac{(\lambda-\lambda_0)^2}{2\sigma^2}} + d
\end{equation}
where $I(\lambda)$ describes the intensity profile with respect to the wavelength $\lambda$, $A$ is the amplitude of the profile, $\lambda_0$ is the Doppler shift, $\sigma$ is the standard deviation associated with the width of the profile, and $d$ is a constant offset to account for the intensity levels at the far wings. Pixels that were associated with a double-peaked RP were fitted with a double-Gaussian function with one component for an emission line and the other for a superimposed absorption line. The fitted absorption line will mimic non-flaring material that blocks the radiation along the line of sight (LOS) of the telescope. We analyse the fitted emission line, under the assumption that the ribbon fine structure will be in emission. If a pixel belongs to a single-peaked RP cluster, one Gaussian in emission is fitted. The distributions of the fitted parameters are presented in Sect.~\ref{sec:fitting_results} and in Table~\ref{tab:riblet_measurements}.


\section{Results}
\label{sec:results}

\subsection{Imaging overview of riblets}
\label{subsec:riblets_overview}

\begin{figure*}[ht!]
    \sidecaption
    \includegraphics[width=\textwidth]{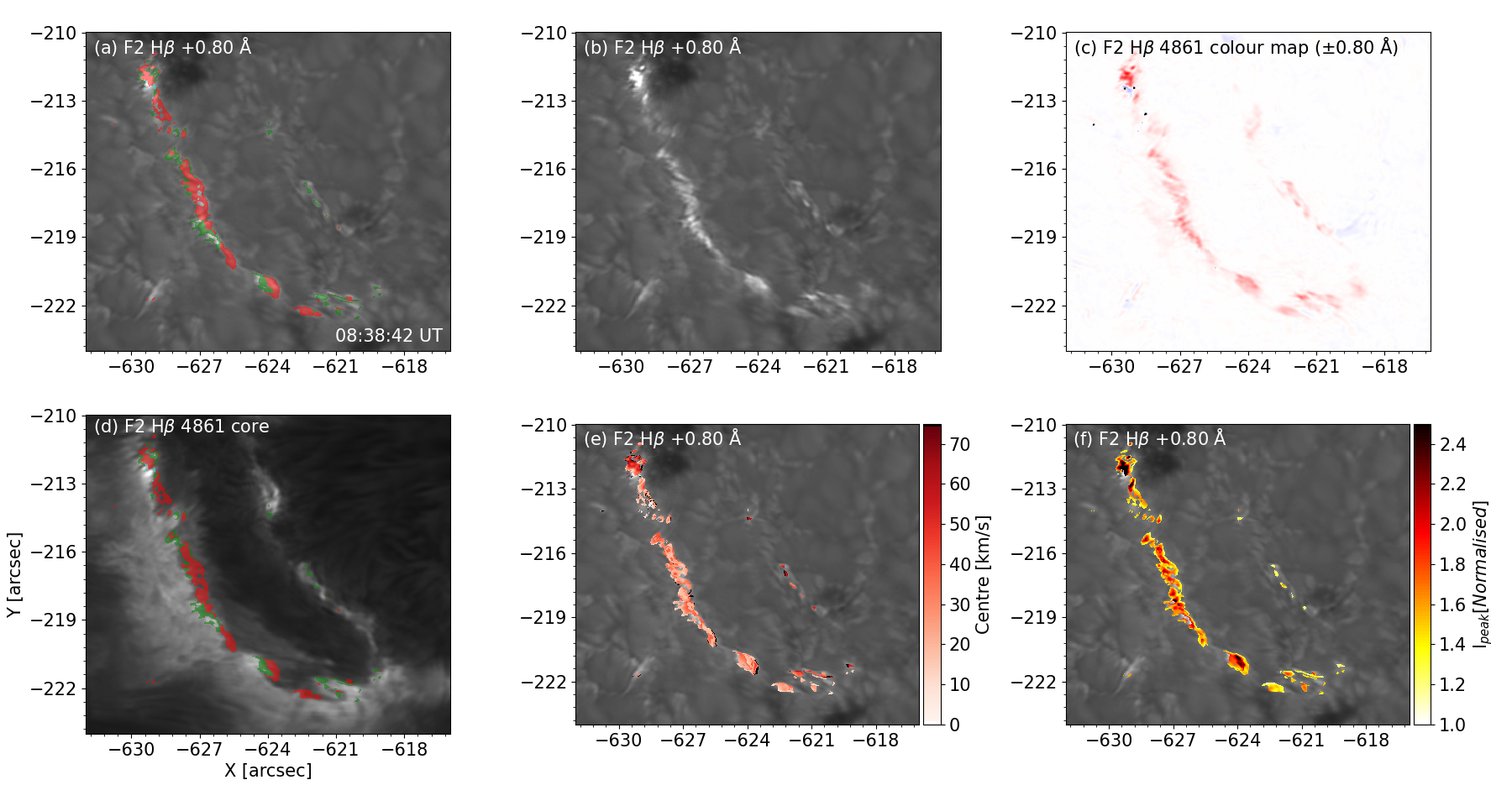}
    \centering
    \caption{
    Same as Fig.~\ref{fig:F1_kmeans_and_Gauss_maps} but for the F2 flare. The superimposed pixels in panel~(f) are replaced with the estimated amplitude $A$ from profile fitting. An animation related to this figure is available online.
    }
    \label{fig:F2_kmeans_and_Gauss_maps}
\end{figure*}

\begin{figure*}[ht!]
    \sidecaption
    \includegraphics[width=\textwidth]{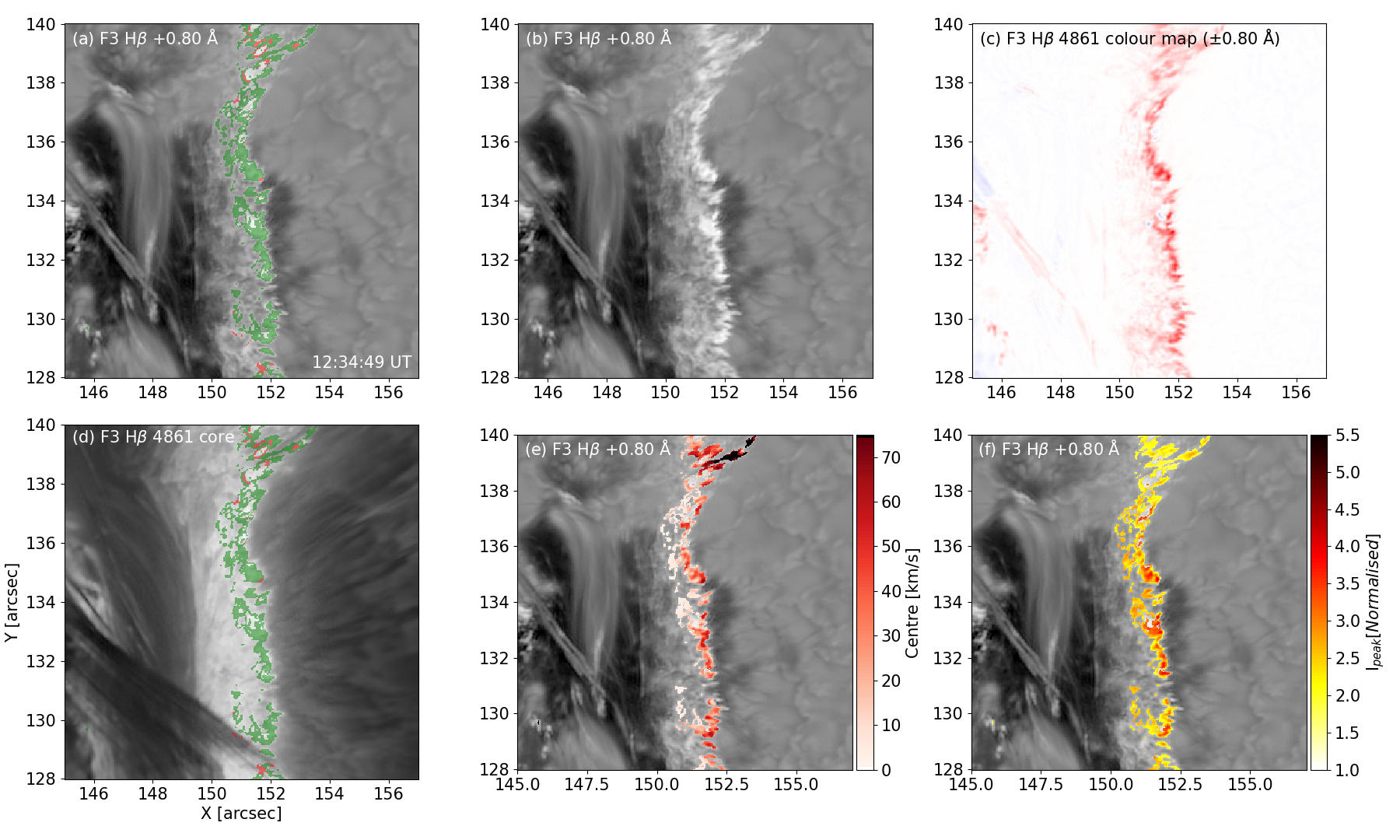}
    \centering
    \caption{
    Same as Fig.~\ref{fig:F1_kmeans_and_Gauss_maps} but for the F3 flare. The overplotted colours in panel~(f) are replaced with the estimated amplitude $A$ from profile fitting. An animation related to this figure is available online.
    }
    \label{fig:F3_kmeans_and_Gauss_maps}
\end{figure*}

The impulsive phase is well covered for the F2 flare and the F3 flare observations, while the F1 flare was only covered from around the GOES peak time. The F2 and F3 flares indicate that the riblets form the ribbon fronts just before the GOES peak time. The collective pattern of riblets propagates along the ribbon during the progression of the flares until they fade. While the F3 flare shows clear signatures of riblets during the impulsive phase and during the peak time, riblets from the F2 flare seem to fade before reaching the GOES peak (see the animation associated with Fig.~\ref{fig:ALL_aia171_context} available in the online material). The F1 riblets are evident near the time of the GOES peak and also show a collective motion along the ribbon. The motion of the ribbon fronts is not discernible from the F1 flare. All three observations show that the riblets jointly constitute the ribbon structure, clearly seen from the H$\beta$~+0.8\AA\ images.

The F1 flare is the only observation close to the limb and therefore provides a side-view into the ribbons. Small-scale brightenings were evident within the ribbons showing the extension of the riblets, seen as bright streaks in Fig.~\ref{fig:F1_kmeans_and_Gauss_maps}b. The riblets evidently constitute an array of features that form the ribbon. The F1 flare indicates that the general structure of riblets is thin streaks that extend vertically upward from lower atmospheric heights, as they typically extend towards the limb. The estimated length and width of the riblets are provided in Table~\ref{tab:riblet_measurements}. Individual features appear to have footpoints that extend further down to a common, continuous structure that stretches parallel to the surface. The continuous structure does not vary much over time compared to the riblets. The larger feature at $[X,Y]=[890\arcsec, 427\arcsec]$ does not exhibit a typical riblet structure, as its width is approximately equal to its length. Images at the H$\beta$ core show that this feature could be a composition of smaller features, as seen in Fig.~\ref{fig:F1_kmeans_and_Gauss_maps}d. The composition is supported by the colour map in Fig.~\ref{fig:F1_kmeans_and_Gauss_maps}c, as smaller confined features are more prominent in this map.

The \kmeans\ clustering effectively identified pixels that were associated with a riblet along the ribbons. All pixels categorised as a riblet were labelled accordingly and overplotted in Fig.~\ref{fig:F1_kmeans_and_Gauss_maps}a and Fig.~\ref{fig:F1_kmeans_and_Gauss_maps}d in colours for the limb flare. The colours associate the pixels with RPs that consist of different profile characteristics, as explained in Sect.~\ref{subsec:identifying_riblets}. The riblets show an overrepresentation of red-dominant double-peaked profiles. \Hbeta~core images reveal that the central reversal of the profiles can be affected by the canopy blocking the LOS in the chromosphere. Symmetrical double-peaked profiles are less frequent and are located only at the riblet footpoints. The larger feature at $[X,Y]=[890\arcsec,427\arcsec]$ is mainly composed of single-peaked emission profiles and is not shadowed by the canopy, as can be seen from the \Hbeta~core in Fig.~\ref{fig:F1_kmeans_and_Gauss_maps}d. The clustering by \kmeans\ effectively tags the riblets along the ribbon, depending on the spectral characteristics within the local area.

The fitted Doppler shifts and profile widths obtained from the main emission component are overplotted in Figs.~\ref{fig:F1_kmeans_and_Gauss_maps}e--f, respectively. The Doppler velocities at the body of the riblets are consistently higher than the footpoints. A thorough investigation of the blue-shifted pixels was deemed unreliable, as the fitted profiles deviate too much from the data. These pixels exhibit a correspondence with a stronger offset in the blue wing at \Hbeta~-0.8~\AA, as can be seen from the colour map in Fig.~\ref{fig:F1_kmeans_and_Gauss_maps}c. The widths of the profiles, as seen in Fig.~\ref{fig:F1_kmeans_and_Gauss_maps}f, suggest that the riblet bodies consist of narrow emission profiles while the footpoints have comparably broad profiles.

The weaker F2 flare exhibits less prominent riblet signatures, although they were detected as part of the ribbon structure. The results from \kmeans\ showed that about 40~\% of the pixels labelled as a riblet were categorised as a single-peaked emission profile, and about 60~\% are related to a double-peaked profile. The location of the riblets is highlighted by the coloured pixels in Figs.~\ref{fig:F2_kmeans_and_Gauss_maps}a and \ref{fig:F2_kmeans_and_Gauss_maps}d, where the latter shows that the riblets are part of the ribbon front. The coloured pixels show that both single-peaked and double-peaked profiles are evident along the ribbon front. This combination may be caused by projection effects, as the conjugate ribbon is located closer toward the limb, causing some of the pixels to be shadowed by material in the LOS. The estimated Doppler velocities are consistently towards the red. Stronger red-shifts are located near regions where signatures of enhanced photospheric magnetic fields are evident, seen as pores, for example at $[X,Y] = [-628\arcsec,-212\arcsec]$ and $[X,Y] = [-620\arcsec,-224\arcsec]$ in Fig.~\ref{fig:F2_kmeans_and_Gauss_maps}e. The estimated profile amplitudes are also more significant closer to the pores.

The results of the \kmeans\ algorithm applied on the F3 DC flare show that the majority of riblets are related to single-peaked RPs, as overplotted in green in Figs.~\ref{fig:F3_kmeans_and_Gauss_maps}a and \ref{fig:F3_kmeans_and_Gauss_maps}d. The colored pixels are located along the ribbon front that is propagating to the west. Double-peaked profiles are evident in areas where chromospheric material is enveloping the ribbon, and can be seen as red pixels at $[X,Y] = [152\arcsec,128\arcsec]$ in Fig.~\ref{fig:F3_kmeans_and_Gauss_maps}d. The parameters from the fitting indicate that stronger red-shifts, $\lambda_0$, are located at the ribbon fronts, similar to those in the F2 flare. The estimated Doppler velocity range is shown in Table~\ref{tab:riblet_measurements}. The strongest profiles in terms of amplitude $A$ are located above pores in the photosphere, as seen by the dark underlying structure in Fig.~\ref{fig:F3_kmeans_and_Gauss_maps}f. The pores are associated with strong negative magnetic fields close to an opposite and near equally strong region to the east, as seen near $[X,Y]=[152\arcsec , 133\arcsec]$ in Fig.~\ref{fig:ALL_aia171_context}f.

\begin{figure*}[ht!]
    \includegraphics[width=\textwidth]{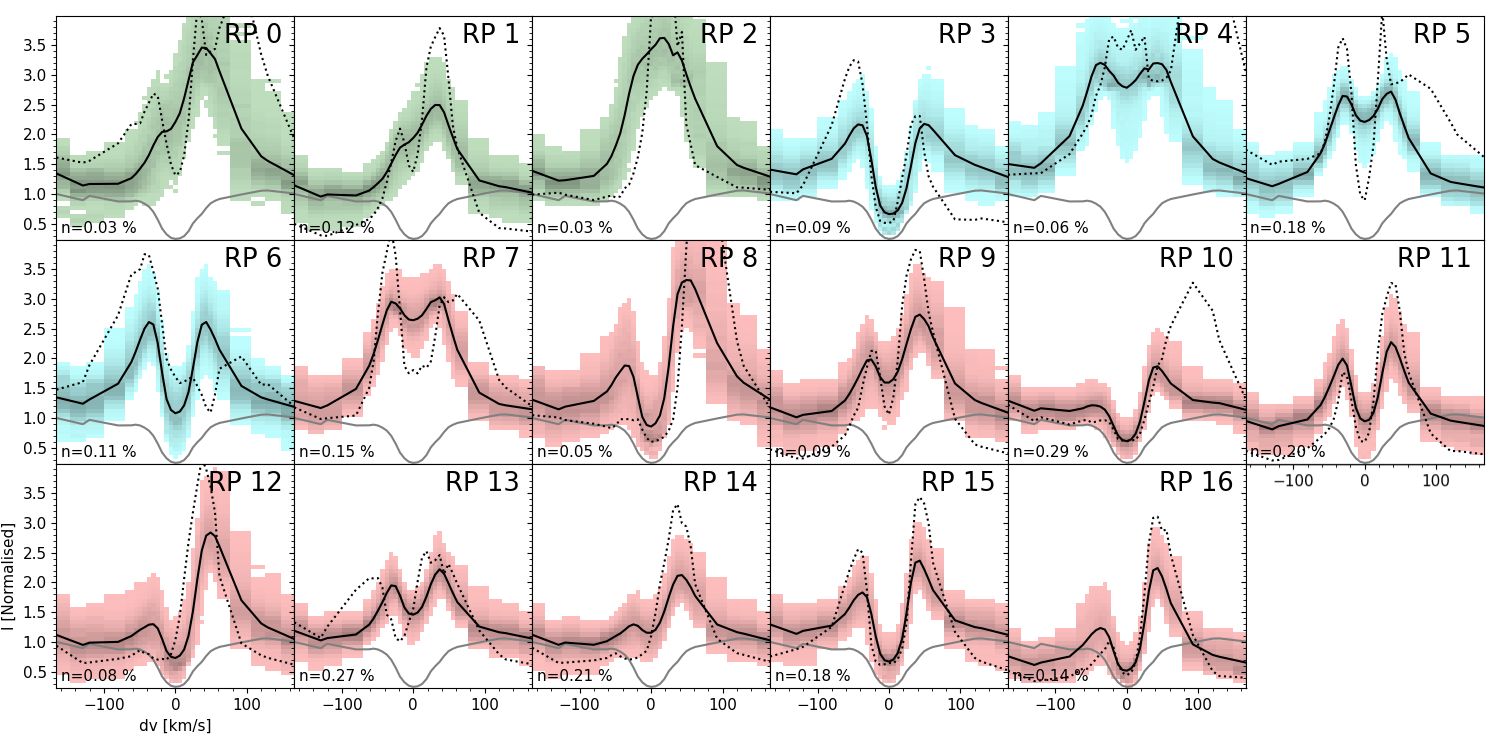}
    \centering
    \caption{
    \Hbeta\ spectral profiles associated with riblets from \kmeans\ clustering of the F1 flare. Solid black lines show the RPs with the identity labelled in the upper right corner, and the coloured background shows the density distribution of all the clustered profiles. Denser distribution is shown in a darker colour. The black dotted line is the profile furthest away from the RP in Euclidean distance. For reference, the grey solid line is the average profile from a QS region. All profiles are normalised by the far blue wing of the average QS profile. The green distributions are associated with apparent single-peaked RPs, cyan distributions are associated with double-peaked and near-symmetric RPs, and the red distributions are associated with double-peaked RPs with a stronger red peak. The parameter $n$ in all panels represents the fraction of each cluster to the total number of pixels in all timeframes used for clustering. The total number of masked pixels used for clustering was $\sim$$7.8 \times 10^{6}$. 
    }
    \label{fig:F1_RPs}
\end{figure*}

\subsection{Spectral characteristics of riblets}

\begin{figure*}[ht!]
    \includegraphics[width=0.95\textwidth]{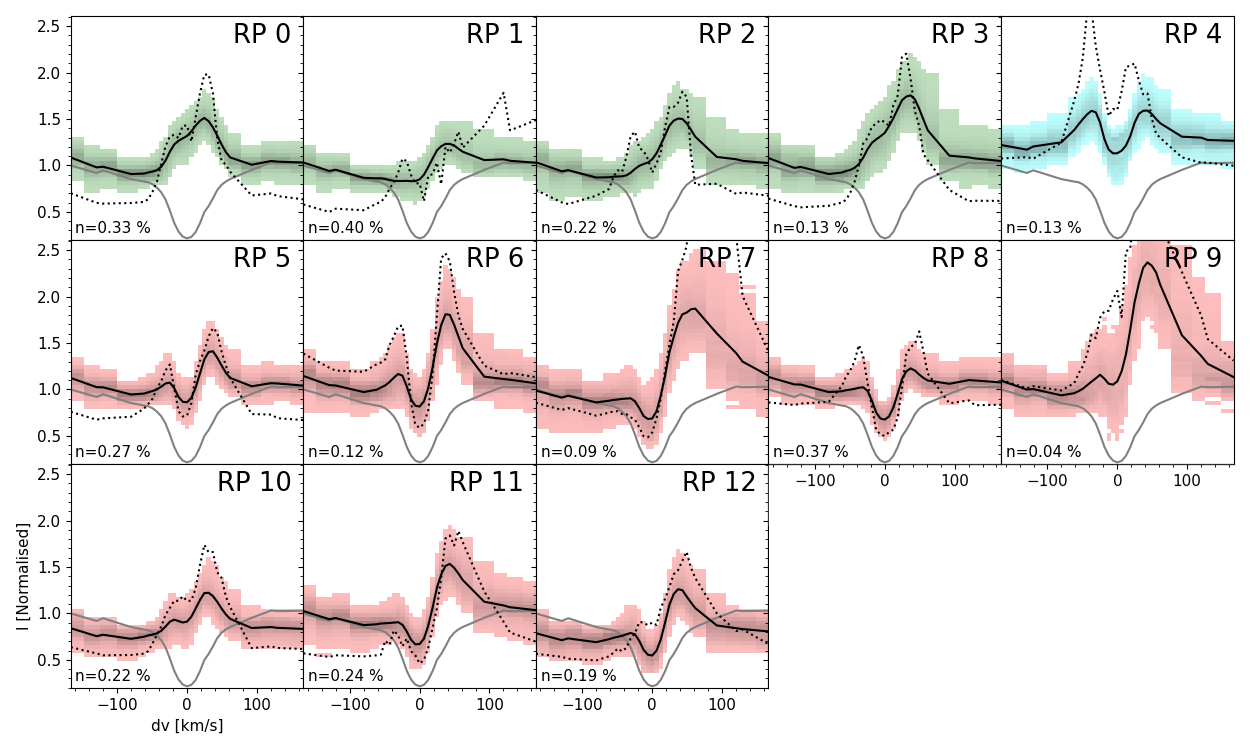}
    \centering
    \caption{
    Same as Fig.~\ref{fig:F1_RPs}, but for the F2 flare.
    A total of $\sim$$6.8 \times 10^{6}$ 
    pixels were used for the \kmeans\ clustering.
    }
    \label{fig:F2_RPs}
\end{figure*}

\begin{figure*}[ht!]
    \sidecaption
    \includegraphics[width=\textwidth]{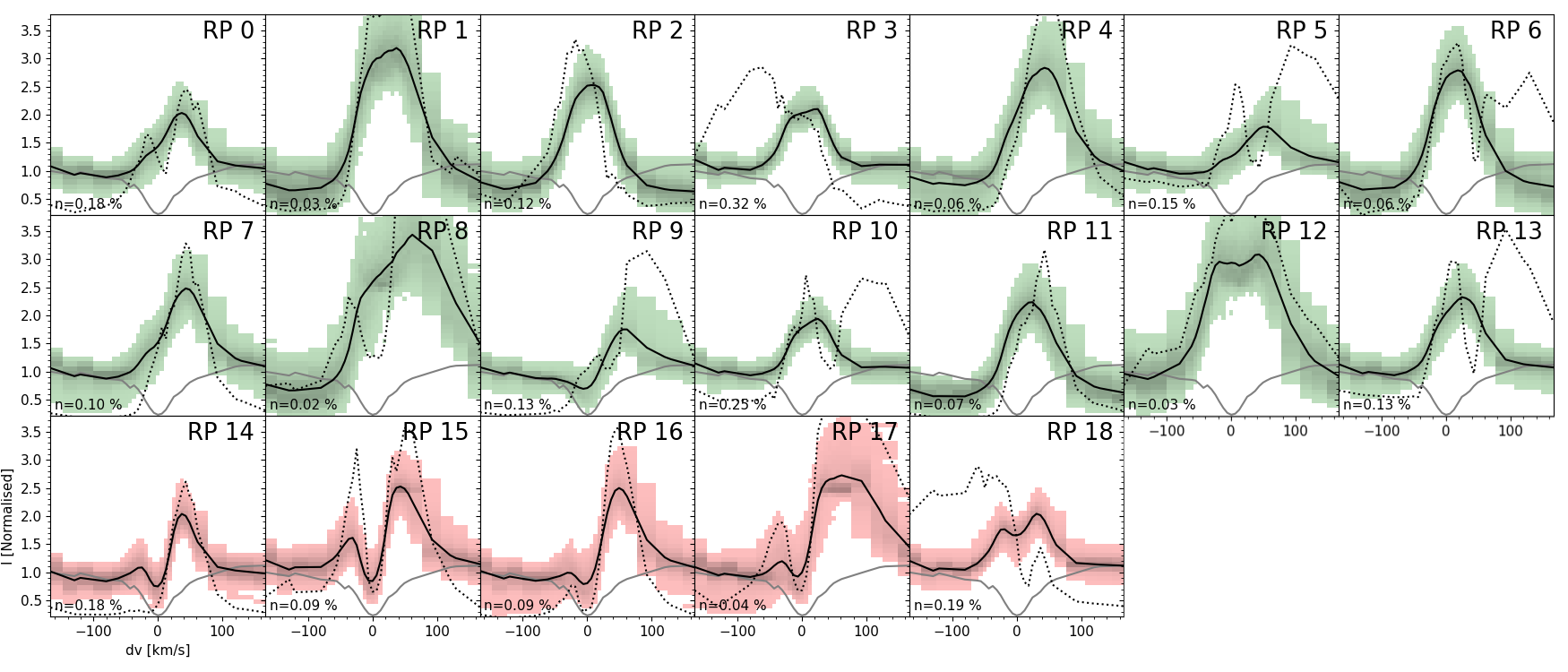}
    \centering
    \caption{
    Same as Fig.~\ref{fig:F1_RPs}, but for the F3 flare. 
    A total of $\sim$$12.3 \times 10^{6}$ 
    pixels were used for the \kmeans\ clustering.
    }
    \label{fig:F3_RPs}
\end{figure*}

The \kmeans\ effectively clustered the spectral profiles into categorical shapes for each dataset. The observations captured the three flares at different viewing angles $\mu$, which provides an augmented analysis. The spectral properties of the riblets show a clear signature in emission, but different compositions of the spectral properties were discerned. Here we show the results of the RPs obtained from the \kmeans\ clustering, and present the spectral characteristics from the three observational datasets.

The RPs obtained from the \kmeans\ clustering of the F1 limb flare generated mainly double-peaked RPs. The RPs from the F1 flare riblets are shown in Fig.~\ref{fig:F1_RPs}. The coloured distribution in each panel is well-aligned with the associated RP. The single-peaked RP~0--2 peaks are red-shifted with evident broadening of the profiles. The green density distributions show that there is a systematic redshift of basically all the profiles in these clusters. The structures of RP~3--6 are nearly symmetric with excessive broadening. A significant RP wing enhancement on either side is not evident. RP~7--16 represent double-peaked profiles with a stronger red peak, which are the most common types for the limb observation. Excessive profile broadening is common for these RPs. RPs with prominent enhanced red wings are common except for RP~7, RP~11 and RP~15. A clear indication from the RPs and the associated observed profiles shows that riblets consist of an emission component.

RPs associated with the F2 flare riblets are generally weaker than the RPs obtained from the F1 flare. The peak amplitudes are typically near or below $A=2$ in normalised units, except for RP~9, which has an amplitude of 2.4, seen in Fig.~\ref{fig:F2_RPs}. Some of the individual profiles in this cluster have a peak amplitude above 2.7. The distribution of pixels within single-peaked RP clusters aligns well with the corresponding RP, and both suggest red-shifts, as shown by the black line and the green distribution, respectively, in RP~0--3. Only one near-symmetric and double-peaked RP was obtained (RP~4). RP~5--12 were associated with double-peak profiles, with a stronger red peak. There is no signature of a prominent blue peak in any of RP~5--12. The red peak is significantly stronger than the blue peak for RPs 6, 7, 9, 11, and 12. The RPs obtained from the F2 flare show that riblets closer to DC ($\mu=0.79$) compared to the F1 flare, are still significantly prone to central reversal signatures.

The F3 DC flare observation has more single-peaked RPs than the other two flares, see Fig.~\ref{fig:F3_RPs}. All the RPs show a red-shifted trend with variable profile amplitudes. The density distribution fits well with the RPs. The variation of the distributions seems to be stronger near the peaks of the RPs and more prominent for the stronger profiles, as seen by comparing RP~3 with RP~8. RP~14--18 has a double-peaked signature with a stronger red peak, although the blue peak in RP~16 and RP~17 is small. The shape of all the red colored profiles suggests redshifts. For all RPs, a red wing enhancement is not a consistent feature. Blue wing enhancements can be discerned from the density distributions, for example, from RP~12 and the dotted line in RP~15.

\subsection{Outcome of profile fitting}
\label{sec:fitting_results}

In this section, we present the properties extracted from profile fitting associated with the riblets. Each pixel was fitted to a single or double-Gaussian curve, based on the category it was labelled to by the \kmeans\ algorithm. Note that some single-peaked pixels were categorised as double-peaked from the \kmeans\ clustering, and vice versa. Such false categorisation is likely due to the vast complexity of flaring profiles, suggesting that the number of clusters \textit{k} could be increased. Increasing the number of clusters implies an excessive number of RPs. Although misplacement of pixels is inevitable for the chosen number of clusters \textit{k}, the fitted emission component is still sufficiently representative of the observed profile. At least, for statistical significance purposes, when estimating properties associated with the riblets.

\begin{figure*}[h!]
    \sidecaption
    \includegraphics[width=\textwidth]{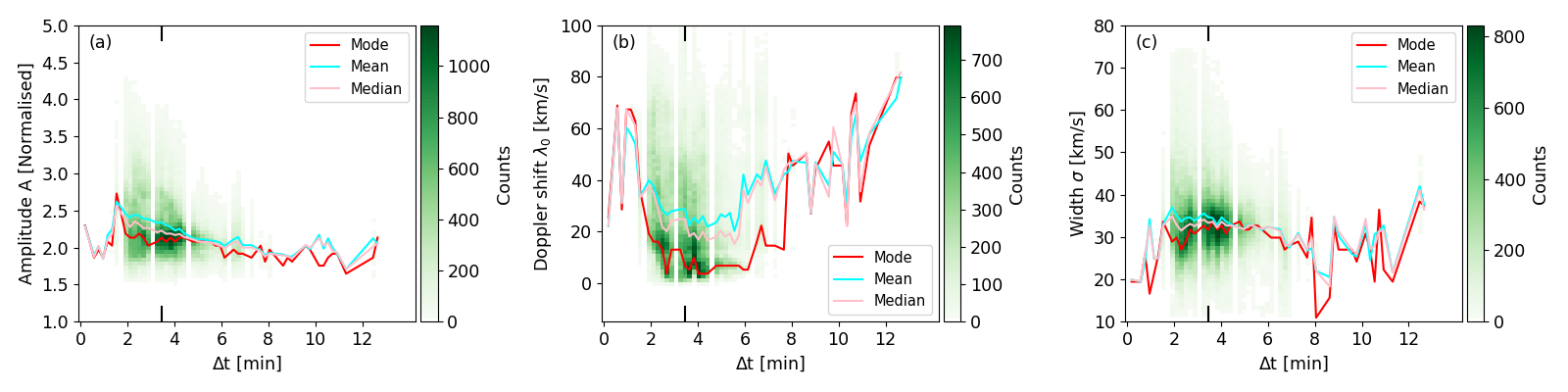}
    \centering
    \caption{F3 riblets evolution of the parameters obtained from the profiles fitted to a single Gaussian curve. The distributions show the fitted amplitude, Doppler shifts and broadening from (a)--(c), respectively. In all panels, the green background represents the distribution with respect to time. The mode, mean and median of each distributions per timestep are shown as red, cyan and pink curve, respectively. The white columns represent frames that were disregarded due to worse seeing. Bins with a contribution less than 1/100 of the maximum bin are coloured in white. The two short vertical black lines at $\Delta t$=3.4~min mark the timestep with the highest number of pixels identified as riblets.
    The distribution of the fitted parameters of this timestep is shown in Fig.~\ref{fig:F3_single_density}.
    }
    \label{fig:F3_all_density}
\end{figure*}

Double-peaked profiles are naturally more complex than single-peaked profiles. By visually inspecting the goodness of the fit on the double-peaked profiles, it was concluded that the procedure struggled in many cases to capture both the main emission component and the central reversal component sufficiently. For instance, instead of capturing the full profile width, the fitting procedure evaluated only the red peak, which yielded inaccurate parameter estimates. This issue led to the exclusion of pixels labelled as double-peaked when interpreting the evolution of the fitted parameters in Fig.~\ref{fig:F3_all_density}.

The amplitude, Doppler shifts and profile widths are determined by the distribution of the fitted parameters, providing a statistical estimate of the properties associated with riblets. The results of the distribution from the F3 flare are presented in Fig.~\ref{fig:F3_all_density}, which shows how the properties evolve over time from a top-view perspective. Only the pixels fitted to a single Gaussian curve are analysed. A clear transition of the increased number of riblets occurs at $\Delta t=2$~min, which is during the impulsive phase. The distribution at this timestep is the most spread, showing peaks above $A=4.0$. The amplitudes of the mode, mean, and median curves are nearly equal to $A\sim 2.5$ at $\Delta t=2$~min. A descending trend is evident for the mean and median curves until the riblets completely fade at around $\Delta t=8$~min, dropping below $A < 2.3$. A descending trend is also evident for the Doppler shift, but for a shorter time range, specifically from $\Delta t$~=2--4.5~min. The mean and median curves are more correlated, starting at $\lambda_0 \sim 35$~\kms\ to $\lambda_0 \sim 25$~\kms. A consistent offset of about $-$~10~\kms\ between the mode curve and the mean/median curve is evident. The highest density of the fitted widths is approximately constant at $\sigma \sim 33$~\kms, except between $\Delta t$~=2--3~min, where the density distributions and the mode curve indicate a trend of increasing widths.

\begin{figure*}[ht!]
    \sidecaption
    \includegraphics[width=\textwidth]{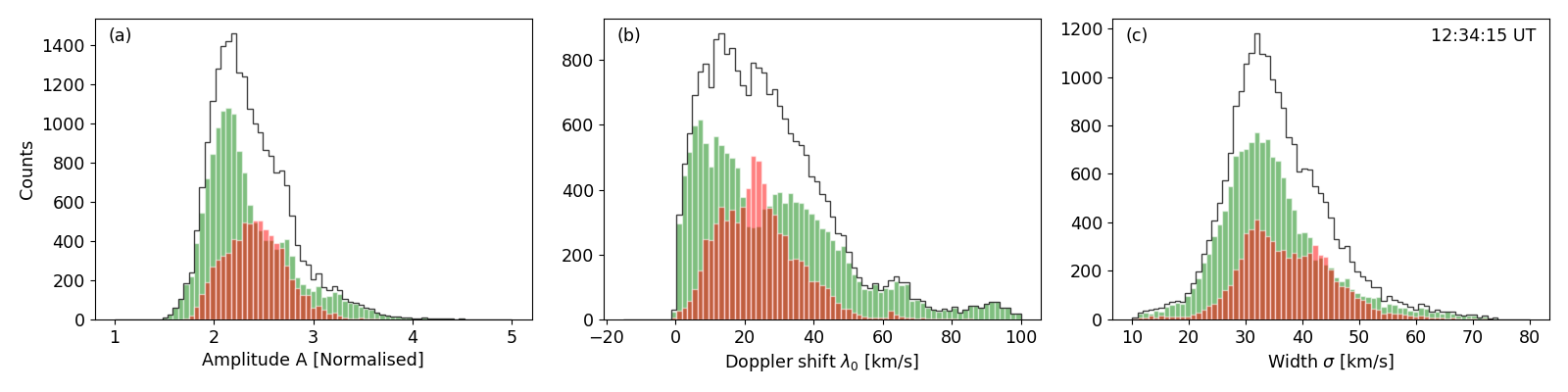}
    \centering
    \caption{Distribution of the fitted parameters obtained from the F3 flare riblets 
    at $\Delta t$=3.4~min just before the GOES peak. This is the timestep with the highest number of pixels identified as riblets. 
    The green bars show the distribution of fitting to a single Gaussian curve. The red bars represent the distribution of parameters obtained from the main emission component of the double-Gaussian fitted profiles. The black curve is the cumulative of the green and red bars.
    }
    \label{fig:F3_single_density}
\end{figure*}

We present the performance of the double-Gaussian fit by comparing its results with the single-Gaussian fit. The results from the F3 DC flare riblets are presented in Fig.~\ref{fig:F3_single_density} at the timestep where most pixels were labelled as a riblet at $\Delta t$~=3.4~min, just before the GOES peak. There is an offset between the peaks of the amplitude distributions, indicating that double-Gaussian profiles are more related to stronger profiles. The Doppler shift distribution for the single-peaked profiles has a valley near 22~\kms, seen by the green bars in Fig.~\ref{fig:F3_single_density}b. The gap shows a correlation with the peak of the double-Gaussian distributions, seen by the red bars in the same panel. The distributions of the widths are fairly similar, except that the distribution for the double-Gaussian fit shows an extra peak at $\sigma \sim 43$~\kms. For all three parameters, the cumulative of both single-Gaussian and double-Gaussian forms a relatively smooth curve, as seen by the black line in all panels in Fig.~\ref{fig:F3_single_density}. The cumulative amplitude and Doppler shift curves show a sharp transition before the curve peak and a wider spread after the peak. The estimated profile widths are nearly symmetric around the peak.

\subsection{Statistical properties of riblets}

\begin{table*}
    \centering 
    \caption{Riblet statistics.}
    \begin{tabular}{c c c c} 
    
    \hline\hline 
     & F1 ($\mu=0.35$, M4.6) & F2 ($\mu=0.79$, C8.3) & F3 ($\mu=0.98$, M1.8) \\ 
    \hline 

    Number of 1(2) peak RPs & 3 (14) & 4 (9) & 14 (5) \\ 

    Fraction of pixels labelled as 1(2) peak & 0.08 (0.92) & 0.40 (0.60) & 0.74 (0.26) \\ 

    Riblet length\tablefootmark{a} & 580--1500~km (0\farcs 79--1\farcs 99) & 280--750~km (0\farcs 38--1\farcs 01) & N/A \\ 

    Riblet width & 110--330~km (0\farcs 15--0\farcs 45) & 80--360~km (0\farcs 11--0\farcs 49) & 110--310~km (0\farcs 15--0\farcs 41) \\ 

    Riblet separation distances & 200--450~km (0\farcs 35--0\farcs 61) & 170--420~km (0\farcs 23--0\farcs 57) & 200--280~km (0\farcs 27--0\farcs 38)   \\

    Fitted amplitude\tablefootmark{b,c} & 2.1--3.8  &  1.2--1.7 & 1.9--2.3 \\ 

    Fitted Doppler shift\tablefootmark{b,c} & 16--32~\kms  &  13--44~\kms & 2--21~\kms \\ 

    Fitted line width\tablefootmark{b,c} & 25--50~\kms  &  12--42~\kms & 25--37~\kms \\ 
    
    \hline 
    
    \label{tab:riblet_measurements}
    \end{tabular}
    \tablefoot{
    \tablefoottext{a}{Projection effects are not accounted for when estimating the length of the riblets.}
    \tablefoottext{b}{The estimated range of the relevant parameter is computed by extracting the maximum and minimum of the FWHM of the distributions. }
    \tablefoottext{c}{For consistency between all three observations, the ranges are obtained from the fitting to a single-Gaussian curve only, as found to be necessary by the $\mu = 0.35$ and $\mu = 0.79$ flares. Additionally, fitted pixels that showed blue shifts were excluded.}
    }
\end{table*}

The statistically significant properties of the riblets are presented in this section from an imaging and a spectral point of view.

Section~\ref{subsec:riblets_overview} presents images of the riblets at different viewing angles, showing that riblets near the limb are prone to be obscured around the \Hbeta\ line core by intervening chromospheric material, unrelated to the flare, along the LOS. This affected the results of the \kmeans\ clustering such that more RPs were categorised as double-peaked, and the number of pixels belonging to this category was favoured. The number of RPs associated with the single-peak or double-peaked category is presented in the first row in Table~\ref{tab:riblet_measurements}. The fraction of riblet pixels that have a single or double peak is presented in the second row. The table shows that a significant number of pixels are categorised as double-Gaussian closer to the limb, with a decreasing trend towards the DC, from 92\%--~26\%. The DC observation is less prone to intervening material, although chromospheric material is periodically flowing above the ribbons, as seen by the dark feature centred at $[X,Y]=[150\arcsec, 130\arcsec]$ in Fig.~\ref{fig:F3_kmeans_and_Gauss_maps}d. Such material flows reflect on the spectral profiles forming a central reversal component and tag the related pixels as double-peaked, as seen by the red pixels in Fig.~\ref{fig:F3_kmeans_and_Gauss_maps}d near $[X,Y]=[152\arcsec, 128\arcsec]$.

We measured the lengths and widths of a selection of prominent riblets from each dataset. The measurements are computed based on the full-width at half-maximum (FWHM) of their spatial intensity profile in the H$\beta$~+0.8~\AA\ wing, which provides a rough estimate of their sizes. We refer to \citetalias{2025A&A...693A...8T} for a detailed explanation of how the measurements are performed. The estimated range of lengths and widths is provided in Table~\ref{tab:riblet_measurements}. Assuming that riblets are vertical relative to the surface and accounting for projection effects, the measured length of 580--1500~km implies a vertical extent of approximately 620--1590~km for the $\mu =0.35$ limb flare. The lengths from the $\mu = 0.79$ flare measured at 280--750~km imply a vertical extent of approximately 460--1220~km, indicating that the riblets are shorter for this flare. We consider that the riblets follow the magnetic field toward coronal heights, implying that the riblets are nearly vertically oriented. We consider two additional cases, assuming an offset of 30 and 45 degrees from the vertical. For the F1 flare, the length ranges are 760--2000~km and 640--1700~km, respectively. For the F2 flare, the estimated riblet length ranges are 280--760~km for both offsets. No lengths for the riblets near DC were measured. The widths are consistent for all three flares.

Measures of the range of values obtained from the spectral profile fitting are presented in the bottom three rows of Table~\ref{tab:riblet_measurements}. We present the maximum and minimum of the FWHM of the parameter distributions. The amplitudes from the three flares show a correlation with the measured X-ray levels from GOES, that is, the F1 flare (M4.6) has a higher range of amplitudes, while the F2 flare (C8.3) has a lower range. The ranges do not overlap between these two observations. The M1.8 flare (F3) has a distribution between these two distributions. The overlapping Doppler velocities between the observations range from 16--21~\kms, but the extremums do not show a clear sign of correlation between the characteristics from each flare. The strongest Doppler shift within the FWHM interval is measured to 44~\kms\ for the $\mu = 0.79$ flare. This flare is associated with the narrowest profiles, as the lower limit of the line width is estimated at 12~\kms. The maximum line width of the DC $\mu = 0.98$ flare is lower than that of the other flares. The acquired ranges presented in Table~\ref{tab:riblet_measurements} suggest a correlation between the fitted amplitudes and the strength of the flare, while a correlation between the strength and the fitted shifts or line widths is not evident. However, any possible correlation needs to be checked with a larger, statistically significant sample.


\section{Discussion}
\label{sec:discussion}


As an extension of our \citetalias{2025A&A...693A...8T}, this detailed study reveals, through colourmapping of the observed flares, that the fine structures form the leading edges of the ribbons. Using \ion{He}{I}~10830~\AA\ \cite{2016ApJ...819...89X} found that ribbon fronts could be as narrow as 340~km wide, which is consistent with our results presented in Table~\ref{tab:riblet_measurements}. \cite{2022ApJ...926..218N} found that the front widths along the ribbon are irregular in space and time, varying from 0.6--1.2~Mm, which is greater than the widths of our riblets. The width of the riblets is indeed irregular along the ribbon, although we found the widths to be less than 600~km. The width of the riblets in \Hbeta~+0.8~\AA\ may not necessarily explain the full width of the ribbon fronts, but rather probe the precise location of the ribbon front footpoints, as the material at the footpoints is more illuminated.

In \citetalias{2025A&A...693A...8T}, we found a distinct periodicity between the blobs with separation distances ranging from 330--550~km. Using similar methods, we found that the distances have an overlapping range of 200--280~km in our current analysis. A plausible evidence of tearing is based on the presence of conjugate ribbons, which are found in all three observations, in addition to the fine-structure that is evident in both conjugates in all three flares. We support the idea that these periodicities of fine structures are related to current sheet dynamics \citep{2025arXiv250400913D} that are possibly due to tearing of small-scale flux ropes \citep{2021ApJ...920..102W}. Discerning similar periodicity was identified from the F1 limb flare, which can be seen in Fig.~\ref{fig:F1_kmeans_and_Gauss_maps}b as the riblets are near equidistant. Although comparable periodicities are evident between the F1 riblets and the blobs in \citetalias{2025A&A...693A...8T}, the variation in sizes between neighbouring riblets is more prominent in the F1 flare. In numerical modelling, the relative sizes between these types of fine structures have not yet been clearly inferred. Given that the riblets are formed due to reconnection in the current sheet, it is not clear if the variation of sizes is caused by irregularities at the reconnection site or if there is significant interaction in the plasma during the transport of energy from the current sheet to the riblets. The latter may have less effect on the weaker flare in \citetalias{2025A&A...693A...8T}, which resulted in the formation of near-equidistant blobs with nearly equal widths.


The number of pixels associated with riblets from all three observations was abundant. Many of the spectral profiles from these pixels showed complex emission properties, which led to relatively unique RPs from the \kmeans\ clustering. A selection of complex profiles is shown as the most distant profiles in all panels in Figs.~\ref {fig:F1_RPs}, \ref{fig:F2_RPs}, and \ref{fig:F3_RPs}. For instance, the distant profile in cluster RP~0 in Fig.~\ref{fig:F1_RPs}, is double-peaked and broadens beyond the spectral window, but was identified in a single-peaked RP cluster. Another example in RP~1 in Fig.~\ref{fig:F2_RPs}, the distant profile seems skewed towards the red, and the blue wing is weaker than the QS average. The distant profile in RP~18 in Fig.~\ref{fig:F3_RPs} shows a blue-shifted emission profile where the peak has a plateau-like shape. These examples of highly complex profiles, among others within the datasets, demonstrate that the fine structures in a flare ribbon generate intricate and diverse emission profiles. Although the frequency of such profile shapes is rare, the most deviant profiles suggest that increasing the number of clusters could be considered to sufficiently capture these profiles and produce RPs that are more similar in shape. On the contrary, analysing a larger number of RP would be time-consuming and inefficient. We argue that it is sufficient to train a \kmeans\ model with $k=100$ and $k=120$ for C-class or M-class flares, respectively.

We categorised the riblets based on the profile shape into two main categories: single-peaked and double-peaked. Both categories were detected in all three flare observations, but the trend indicates that the occurrence rate of single-peaked profiles is increasing from limb to DC. The double-peaked profiles are generally characterised by a stronger red peak. The shifts of the central reversals were $< 5$~\kms\ for the F2 and F3 DC riblets and around $0$~\kms\ for the F1 limb riblets. Assuming the riblets to be the footpoints of magnetic loops and that plasma is accelerated downwards from the current sheet, red-shifted profiles are more prominent when the magnetic fields are aligned with the LOS, which explains why the central reversal is more red-shifted for the F3 DC observation (Compare, for example, RP~7--16 in Fig~\ref{fig:F1_RPs} with RP~14--18 in Fig~\ref{fig:F3_RPs}). Since detailed information about the magnetic field geometry is unknown and there is an increased chance of an absorption component at more inclined viewing angles, the question of whether the Doppler offset of the main emission component in the F1 limb profiles can be interpreted as a clean measurement of the plasma velocity remains uncertain. However, we note that red-dominant profiles in the chromosphere are not necessarily a sign of downflows since a blue-shifted absorption component can produce a red-shifted emission profile \citep{2015ApJ...813..125K, 2025ApJ...989..183Y}. Nevertheless, we argue that the observed riblets are part of magnetic loops extending in the corona, consistent with tearing-mode theories \citep{2021ApJ...920..102W, 2025arXiv250400913D}. Hence, we interpret the red-shifts as plasma downflows. Images in \Hbeta~core channel from the three flares reveal that the chance of structures in the chromosphere blocking the LOS of riblets is increasing towards the limb, which causes more double-peaked profiles to be formed. The chromospheric canopy in the F1 limb observation is dense enough to block a significant amount of the radiation from the riblets (see red transparent pixels in Fig.~\ref{fig:F1_kmeans_and_Gauss_maps}d). Only the strongest feature at $[X,Y]=[890\arcsec,427\arcsec]$ penetrates the canopy. The riblets from the F3 flare are mostly associated with single-peaked profiles, as seen by the green transparent pixels in Fig.~\ref{fig:F3_kmeans_and_Gauss_maps}d. Double-peaked profiles in red are indeed evident at $[X,Y]=[152\arcsec,128\arcsec]$ and are blended by chromospheric material overarching the riblets. Therefore, the spectral properties of ribbon fine structures are likely always in full emission and blocked by overlying material when observed with two peaks.

Very recently, \cite{2025ApJ...989..183Y} analysed ribbon fine-scale structures using the Daniel K. Inouye Solar Telescope \citep[DKIST,][]{2020SoPh..295..172R}, which show similar spatial properties of the blobs as in our \citetalias{2025A&A...693A...8T}. They found that the \ion{Ca}{II}~8542~\AA\ lines of the blobs exhibit double-peaked profiles, which is contrary to \citetalias{2025A&A...693A...8T} that only showed single-peaked profiles. The study presented here, based on three flares occurring at different positions on the solar disk, suggests that the formation of double-peaked chromospheric profiles is more likely at larger distances from disk-centre, as exemplified by the limb-near flare in \cite{2025ApJ...989..183Y}.


The fitting of double-peaked profiles was challenging due to the diverse complexity of these profiles. Therefore, these pixels need to be carefully analysed when interpreted, preferably as a standalone analysis of these particular pixels. The fitting of individual pixels associated with a riblet enabled us to construct an overview of the plasma dynamics by overplotting the fitted parameters on top of the images. In Fig.~\ref{fig:F1_kmeans_and_Gauss_maps}e, we show the fitted Doppler velocities of the F1 flare. We emphasise that the riblets are observed with a significant inclination angle, indicating that the velocities perpendicular to the surface may be higher. The body of the riblets is mainly red-shifted. The red-shifts are apparently trending towards $0$~\kms\ closer to the footpoints. A correlation with the line widths can be discerned, as the riblet bodies generally show narrower widths, and the footpoint widths are excessive (Fig.~\ref{fig:F1_kmeans_and_Gauss_maps}f). It appears that the plasma is decelerated along the riblet structures and brakes near the footpoint. This potential velocity gradient seems to be associated with line widths, as the widths are broader at the riblet footpoints.

The riblets from the F2 and F3 flares were observed from a top view, which prevents identifying the body and footpoints of the riblets. The fitted Doppler shifts are seen in Fig.~\ref{fig:F2_kmeans_and_Gauss_maps}e and Fig.~\ref{fig:F3_kmeans_and_Gauss_maps}e, obtained from the F2 and F3 flares, respectively, which show that the riblets are generally covered by red-shifted pixels. Since the footpoints of the riblets are not evident, a change in Doppler velocity from the riblet body to the footpoint is not evident. Both F2 and F3 exhibit a potential correlation between strong profiles and pores in the photosphere. This is seen in Fig.~\ref{fig:F2_kmeans_and_Gauss_maps}f and Fig.~\ref{fig:F3_kmeans_and_Gauss_maps}f for F2 and F3, respectively. The prior shows a patch of dark pixels, representing strong profiles, near $[X,Y]=[-630\arcsec,-213\arcsec]$, which is adjacent to a photospheric pore. A similar signature is observed in the latter figure, near $[X,Y]=[152\arcsec,133\arcsec]$, where strong profiles are overlapping the underlying pores. Pores in the photosphere are formed due to strong magnetic density, which implies that the formation of stronger \Hbeta flare profiles may be associated with regions where underlying strong magnetic fields are present.


\section{Conclusions}
\label{sec:conclusions}

Based on high-resolution SST observations of three solar flares, we conclude that flare ribbons consist of extremely fine structures (or blobs) when viewed close to the solar limb. These structures were identified as riblets, according to the definition first introduced by \cite{2025arXiv250701169S}. Riblets are elongated fine structures that form the ribbon fronts. These features were detected in all three flares, which complements the previous detection of ribbon fine-structure from ground-based observations \citep{2015ApJ...810....4B, 2021ApJ...922..117F, 2025arXiv250701169S} and the more recent observations with DKIST  \citep{2025ApJ...989..183Y}. We suggest that such fine-structure might be common in flare ribbons, but a comprehensive statistical analysis is needed to confidently establish the ubiquity of ribbon fine-structure in all C-class flares and stronger.

In our analysis, it becomes evident that the analysed flare ribbons are composed of arranged small-scale features -- riblets --  and are not a continuous structure. As the ribbons constitute the footpoints of reconnection loops that extend into the corona, the array of fine structures along the ribbon suggests that the reconnection process in the solar corona is fragmented. The fragmentation is likely associated with the formation of magnetic islands in the current sheet, although this has not been directly observed. The ribbon fine structures are observational evidence that supports the theoretical idea of tearing in the flare current sheet or the patchy reconnection. 
    
The spectral shapes of riblets are mainly red-shifted emission profiles. The profile shapes from the blobs in \citetalias{2025A&A...693A...8T} are of a similar nature, with red-shifted emission profiles, and are consistent with this statistical analysis. Double-peaked profiles are likely produced by an absorption component due to the presence of material above the H$\beta$ formation height along the LOS. 

In this way, high-resolution observations are essential for advancing our understanding of fine-scale solar flare dynamics. The recently published catalogue of 19 solar flares observed with the SST by \citet{Wilde2025} offers a valuable resource for such studies. Looking ahead, we anticipate further insights from upcoming observations by DKIST, Solar Orbiter \citep{SolarOrbiter2020}, and the European Solar Telescope \citep[EST;][]{EST2022}.



%
%

\bibliographystyle{aa} 
\bibliography{references} 

\begin{acknowledgements}
    We acknowledge our fruitful discussions with Guillaume Aulanier, Lyndsay Fletcher, and Hugh Hudson in Oslo. 
    We thank Semya Amouche Tønnessen for participation in the 2023 SST campaign. 
    This research is supported by the Research Council of Norway, project number 325491, 
    through its Centres of Excellence scheme, project number 262622, 
    and the European Research Council through the Synergy Grant number 810218 (``The Whole Sun'', ERC-2018-SyG).
    The Swedish 1-m Solar Telescope is operated on the island of La Palma by the institute for Solar Physics of Stockholm University in the Spanish Observatorio del Roque de los Muchachos of the Institutio de Astrofisica de Canarias.
    The Swedish 1-m Solar Telescope, SST, is co-funded by the Swedish Research Council as a national research infrastructure (registration number 4.3-2021-00169).
\end{acknowledgements}

\appendix \label{appendix}
\onecolumn

\section{Evaluating the number of clusters}
\label{app:number_of_clusters}

\begin{figure*}[ht!]
    \sidecaption
    \includegraphics[width=\textwidth]{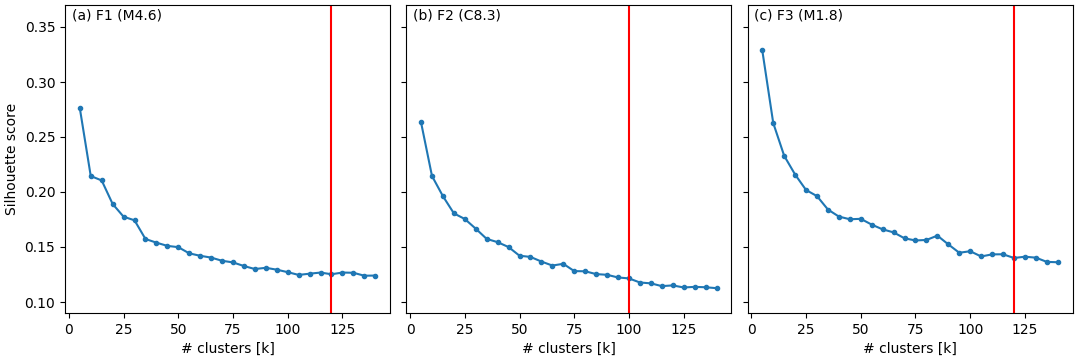}
    \centering
    \caption{
    Silhouette score from the \kmeans\ algorithm. Each panel corresponds to the different flares with an increasing number of clusters, as denoted by the panel titles. For every iteration, a new subset of 25~000 random datapoints was used to compute the scores. The red vertical line in all panels denotes the number of clusters that were chosen for each dataset.
    }
    \label{fig:ALL_silhouette_score}
\end{figure*}

We computed the silhouette score to decide the optimal number of clusters for each of the flares in the analysis \citep{ROUSSEEUW198753}. The silhouette score ranges from -1 to 1. A score of 0 implies that the datapoints are located at the boundary between two clusters. A score close to 1 implies that the datapoints are located relatively distant from neighbouring clusters, while a score close to -1 implies that the datapoint may have a better fit in another cluster. In our analysis, many spectral lines are relatively similar. The similarities between spectral lines are projected in the score plot in the sense that the silhouette score is consistently decreasing with increasing value of \textit{k}, as seen in Fig.~\ref{fig:ALL_silhouette_score}. Therefore, we conclude that the optimal silhouette score is where the curve flattens, which occurs approximately at $k = 100$, with scores of 0.13, 0.12, and 0.15 for the F1, F2, and F3 flares, respectively. See the red vertical line in Fig.~\ref{fig:ALL_silhouette_score}b. The next step involved checking if the clustering effectively captured the pixels associated with a riblet. This process was done by iterating through the clusters and highlighting the related pixels on the \Hbeta+0.8~\AA\ images. The clusters highlighting a riblet were identified. We noted that during this step for the F1 and F3 flare, some riblet-related clusters included a significant amount of pixels that were not considered as riblets. Additionally, when inspecting the most distant datapoint from the cluster centroid in terms of Euclidean distance, more variation in the profile shapes was evident, which may exclude profiles of interest. We approached this issue by increasing the number of clusters to $k = 120$ for these two flares, which improved the accuracy of clustering the riblet-related pixels.

\section{Spectral properties from limb and low viewing angle observation}
\label{app:spectral_properties}

The clustering of the F1 and F2 flare pixels resulted in an abundant representation of double-peaked profiles. We concluded in Sect.~\ref{sec:fitting_results} that the fitting procedure did not reliably capture the main emission component and the central reversal component in many cases. Therefore, we do not include the parameter distribution figures in the main body of the article. For completion, we present the figures in the appendix.

\begin{figure*}[ht!]
    \sidecaption
    \includegraphics[width=\textwidth]{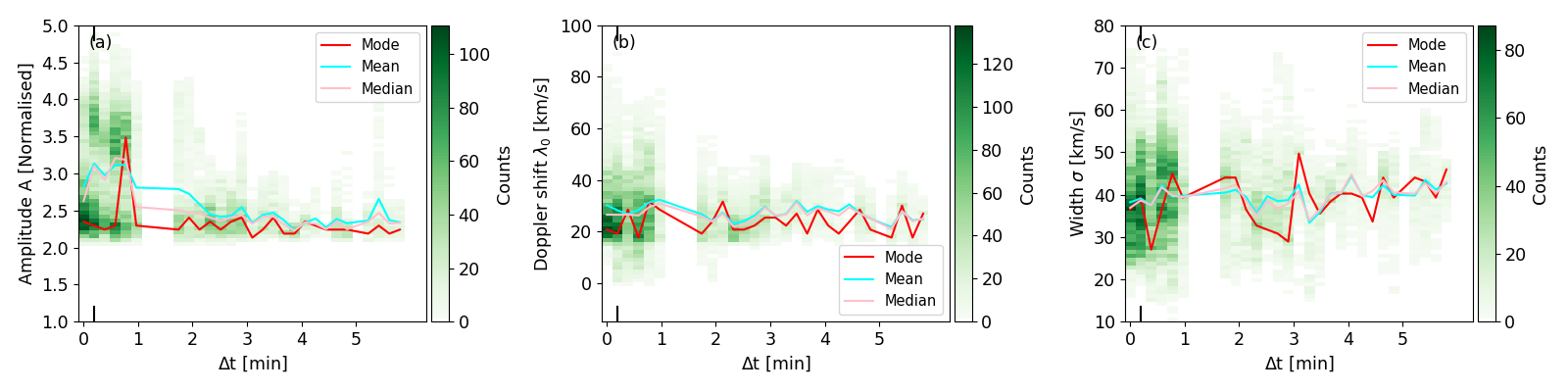}
    \centering
    \caption{
    Same as Fig.~\ref{fig:F3_all_density}, but for the F1 ($\mu = 0.35$) flare. The figure show the results only from pixels fitted to a single Gaussian curve.
    }
    \label{fig:F1_all_density}
\end{figure*}

\begin{figure*}[ht!]
    \sidecaption
    \includegraphics[width=\textwidth]{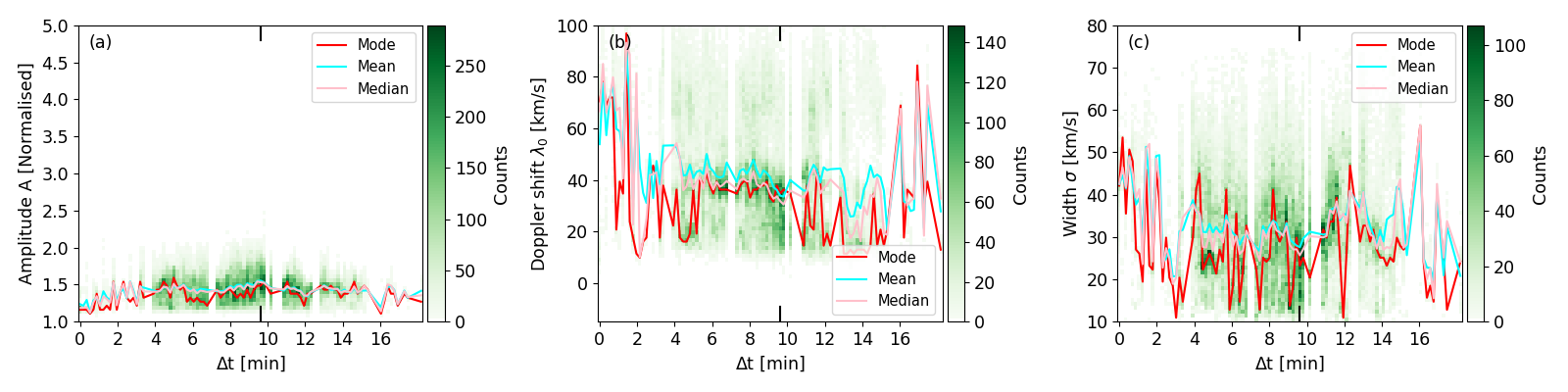}
    \centering
    \caption{
    Same as Fig.~\ref{fig:F3_all_density}, but for the F2 ($\mu = 0.79$) flare. The results include only pixels that were fitted to a single Gaussian curve.
    }
    \label{fig:F2_all_density}
\end{figure*}

\begin{figure*}[ht!]
    \sidecaption
    \includegraphics[width=\textwidth]{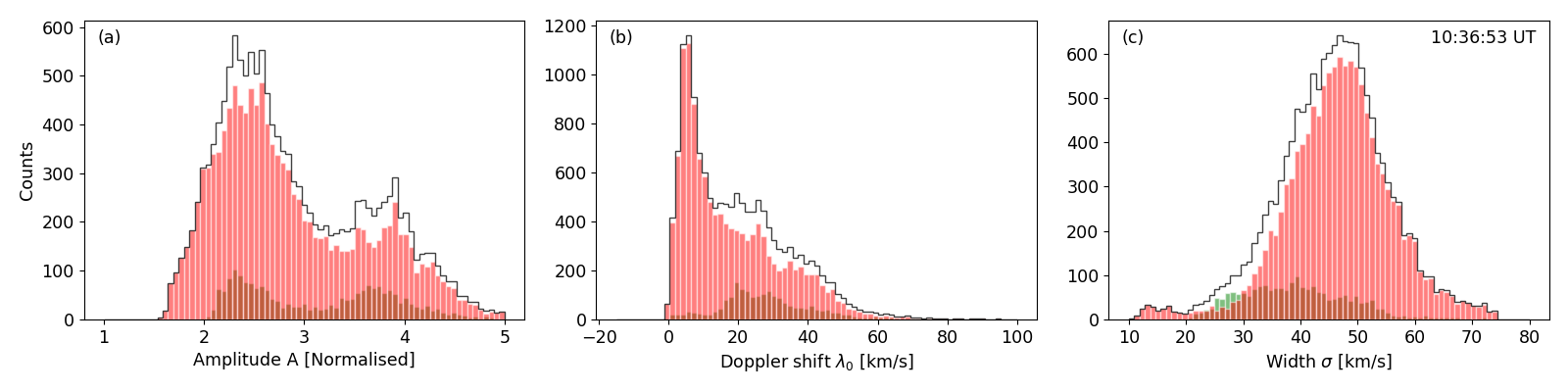}
    \centering
    \caption{
    Same as Fig.~\ref{fig:F3_single_density}, but for the F1 ($\mu = 0.35$) flare. The results include only pixels that were fitted to a single Gaussian curve.
    }
    \label{fig:F1_single_density}
\end{figure*}

\begin{figure*}[ht!]
    \sidecaption
    \includegraphics[width=\textwidth]{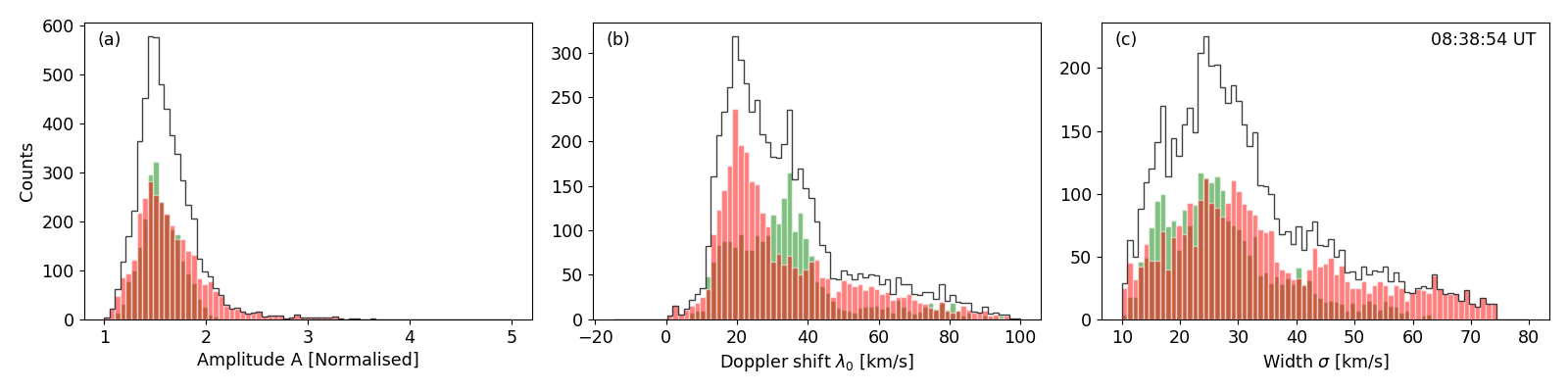}
    \centering
    \caption{
    Same as Fig.~\ref{fig:F3_single_density}, but for the F2 ($\mu = 0.79$) flare.
    }
    \label{fig:F2_single_density}
\end{figure*}


\end{document}